\documentclass[aps,pra,showpacs,column,twocolumn]{revtex4-1}

\usepackage{graphicx}
\usepackage{mathrsfs}
\usepackage{bbm}
\usepackage{amssymb}
\usepackage{amsmath}
\usepackage{subfigure}
\usepackage{slashbox}
\usepackage{bm}
\usepackage{txfonts}

\newcommand{\ket}[1]{|#1\rangle}
\newcommand{\bra}[1]{\langle#1|}

\newcommand{\lr}\longrightarrow
\newcommand{\ra}\rightarrow

\newcommand{\ua}{\uparrow}
\newcommand{\da}{\downarrow}

\newcommand{\Ei}{\raisebox{-0.3\height}{\includegraphics[height=1cm]{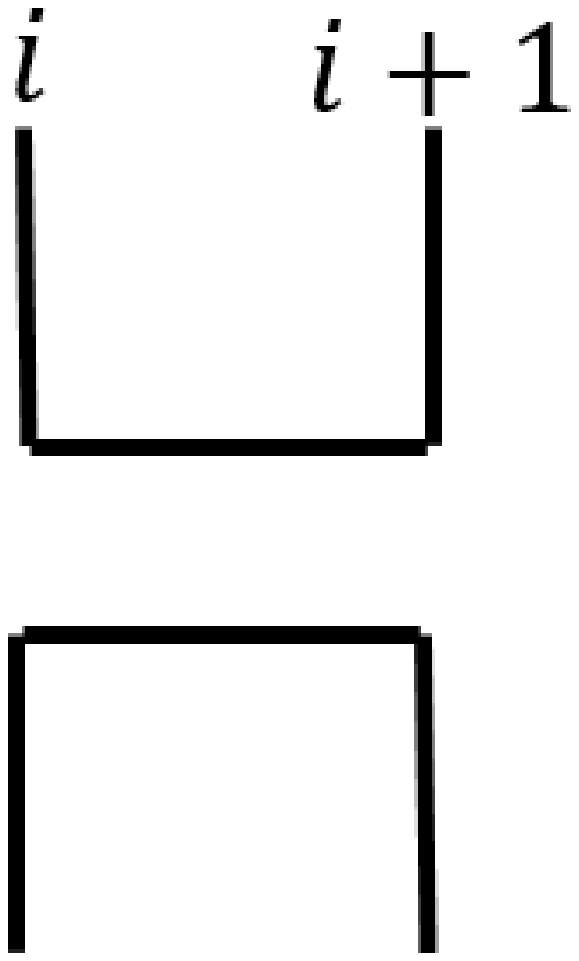}}}
\newcommand{\Bi}{\raisebox{-0.3\height}{\includegraphics[height=1cm]{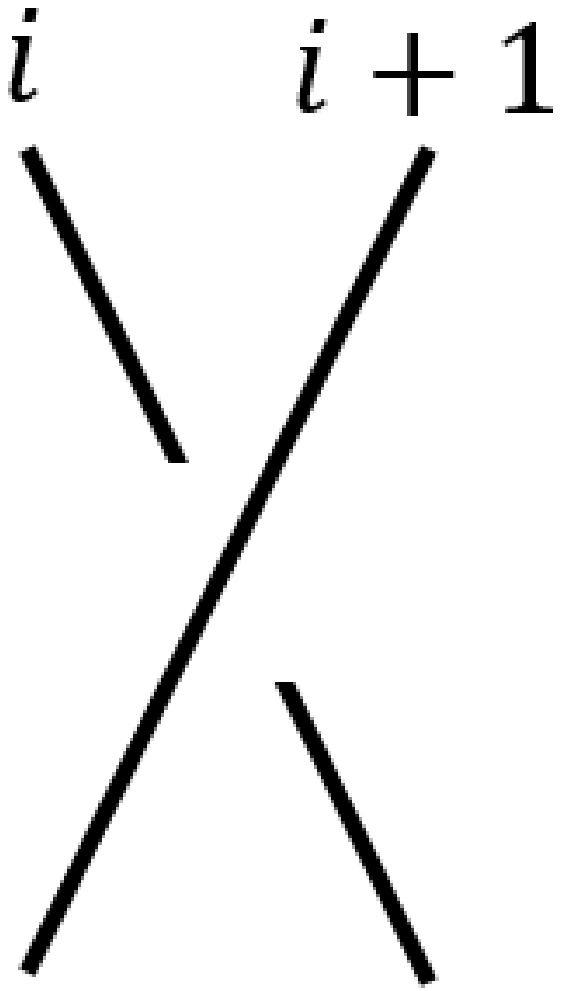}}}
\newcommand{\Bim}{\raisebox{-0.3\height}{\includegraphics[height=1cm]{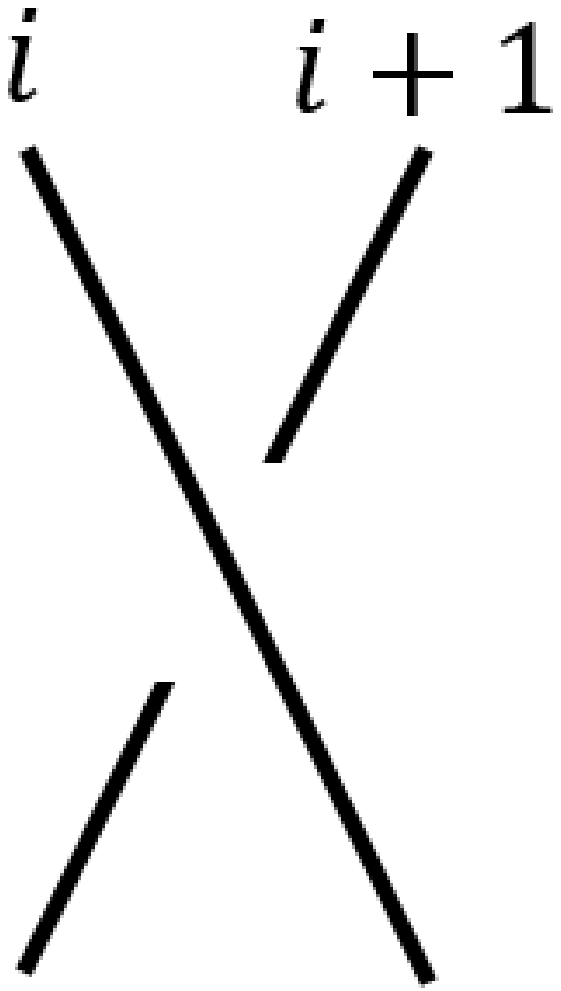}}}

\newcommand{\gx}{\raisebox{-0.3\height}{\includegraphics[height=0.6cm]{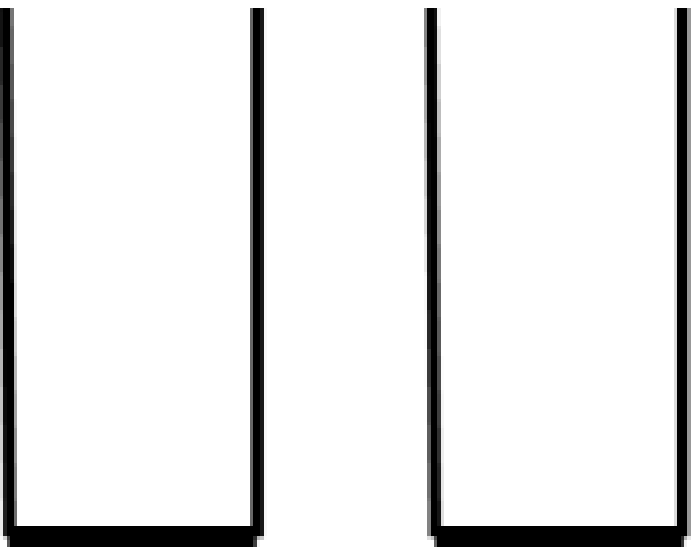}}}
\newcommand{\gy}{\raisebox{-0.3\height}{\includegraphics[height=0.6cm]{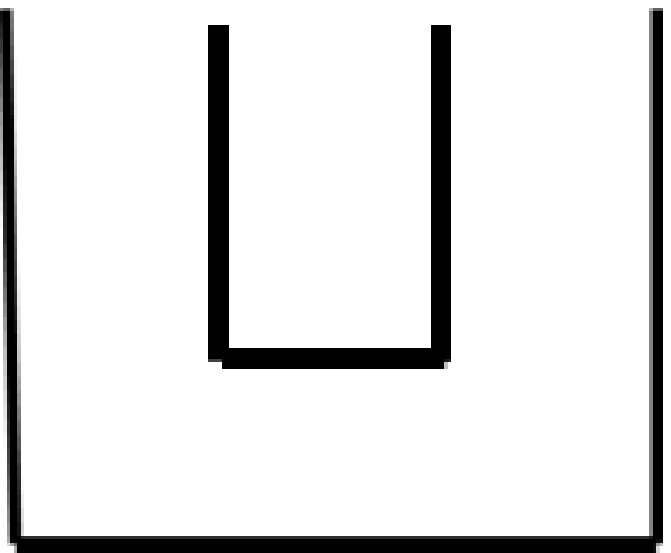}}}
\newcommand{\gz}{\raisebox{-0.3\height}{\includegraphics[height=0.6cm]{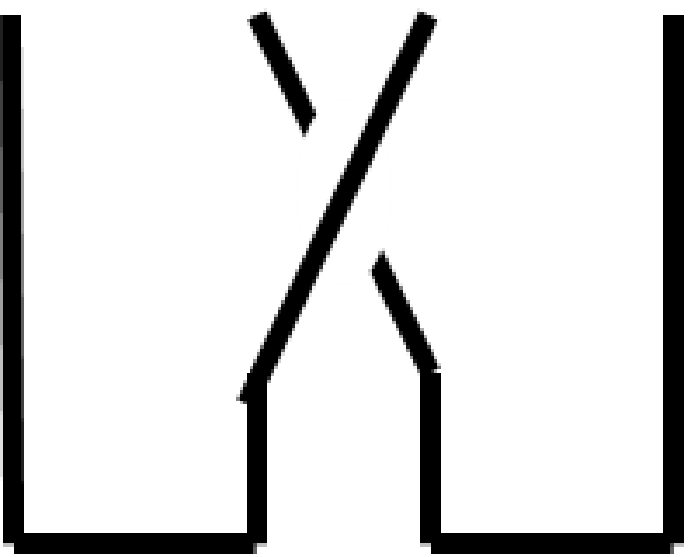}}}

\newcommand{\loopx}{\raisebox{-0.3\height}{\includegraphics[height=0.6cm]{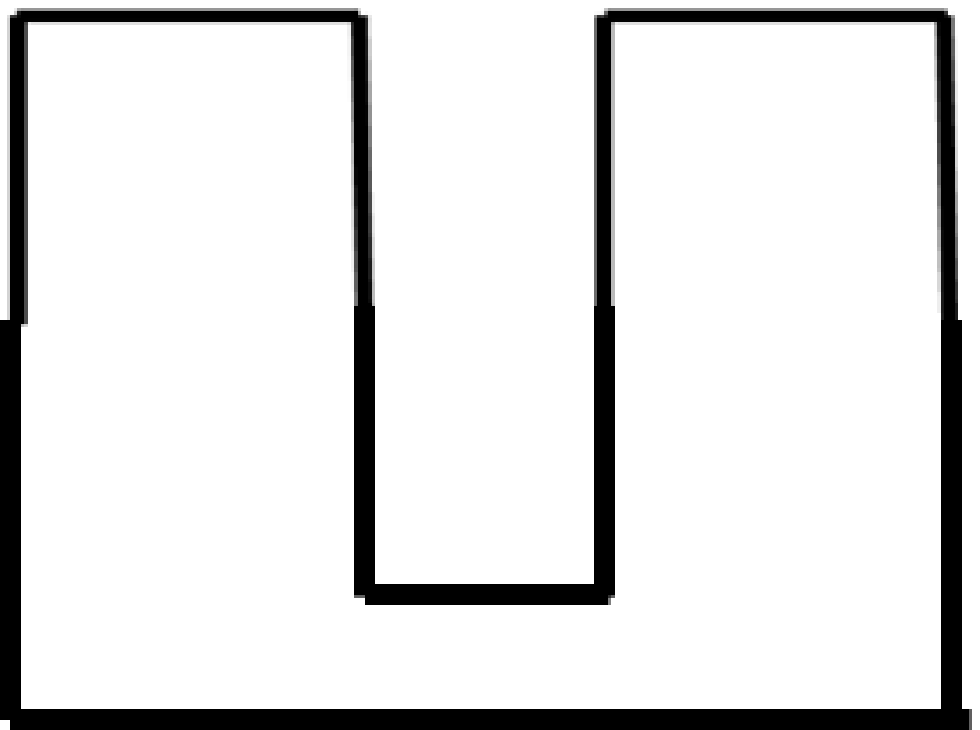}}}
\newcommand{\loopxt}{\raisebox{-0.3\height}{\includegraphics[height=0.6cm]{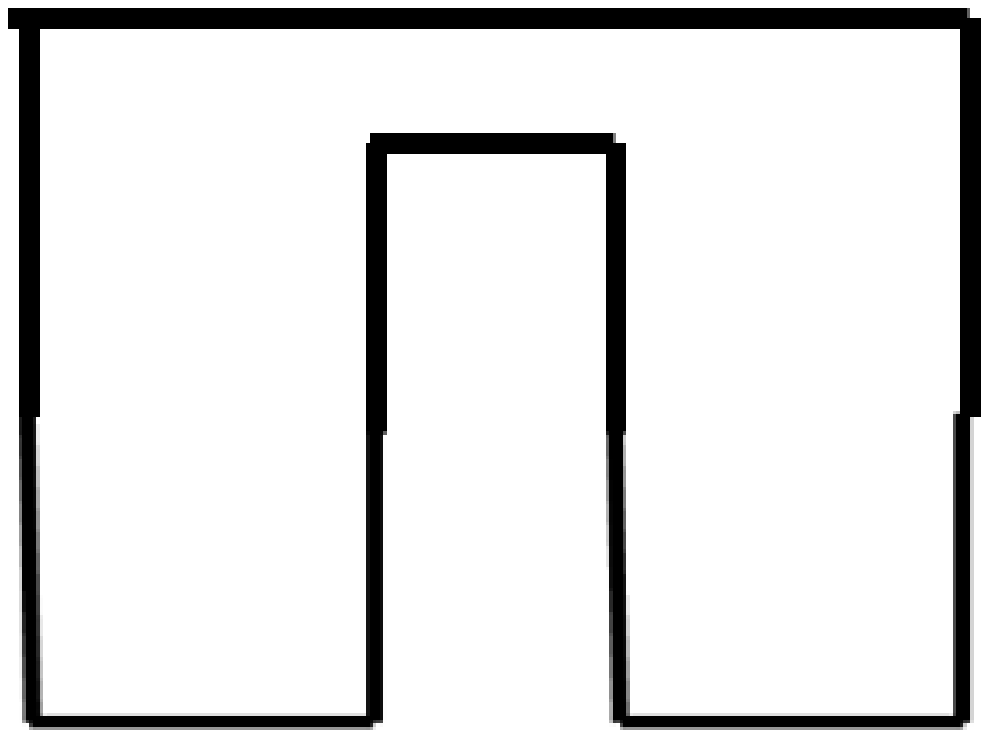}}}
\newcommand{\loopy}{\raisebox{-0.3\height}{\includegraphics[height=0.6cm]{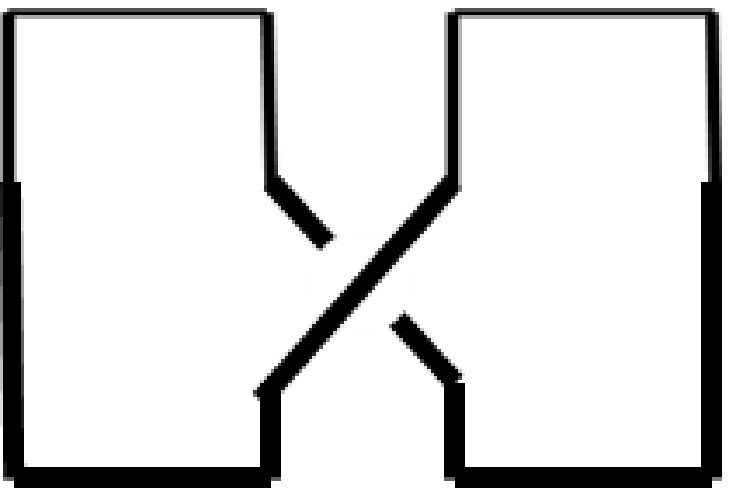}}}
\newcommand{\loopz}{\raisebox{-0.3\height}{\includegraphics[height=0.6cm]{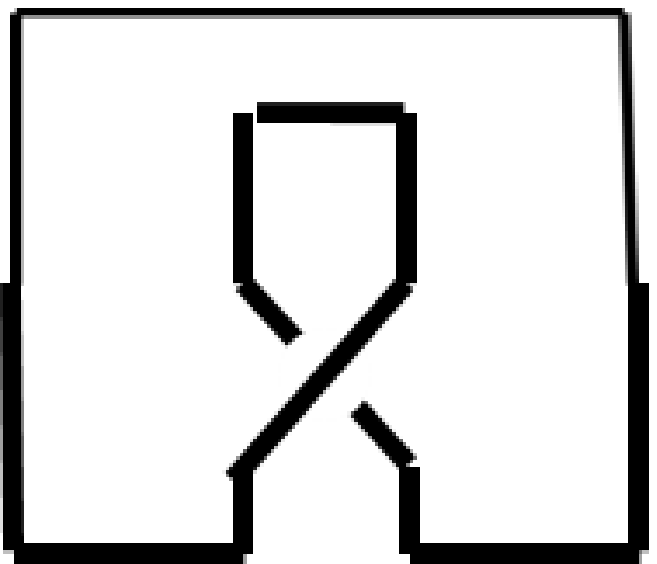}}}
\newcommand{\loopa}{\raisebox{-0.3\height}{\includegraphics[height=0.6cm]{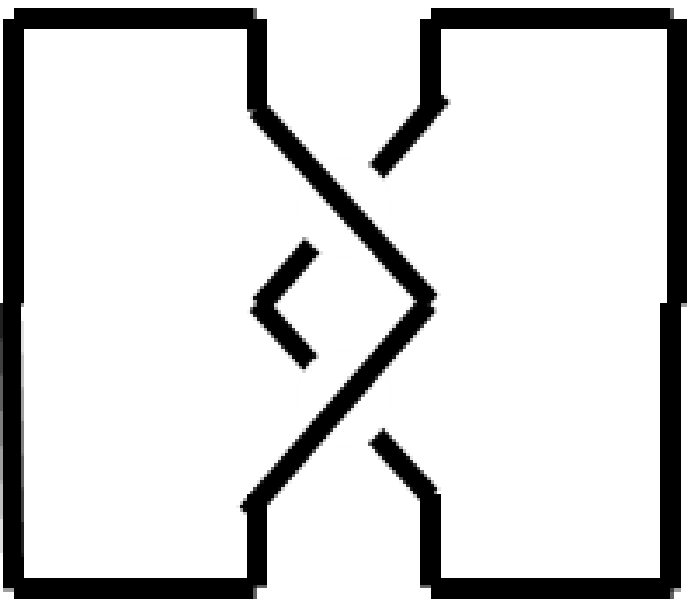}}}

\newcommand{\ta}{\raisebox{-0.3\height}{\includegraphics[height=0.6cm]{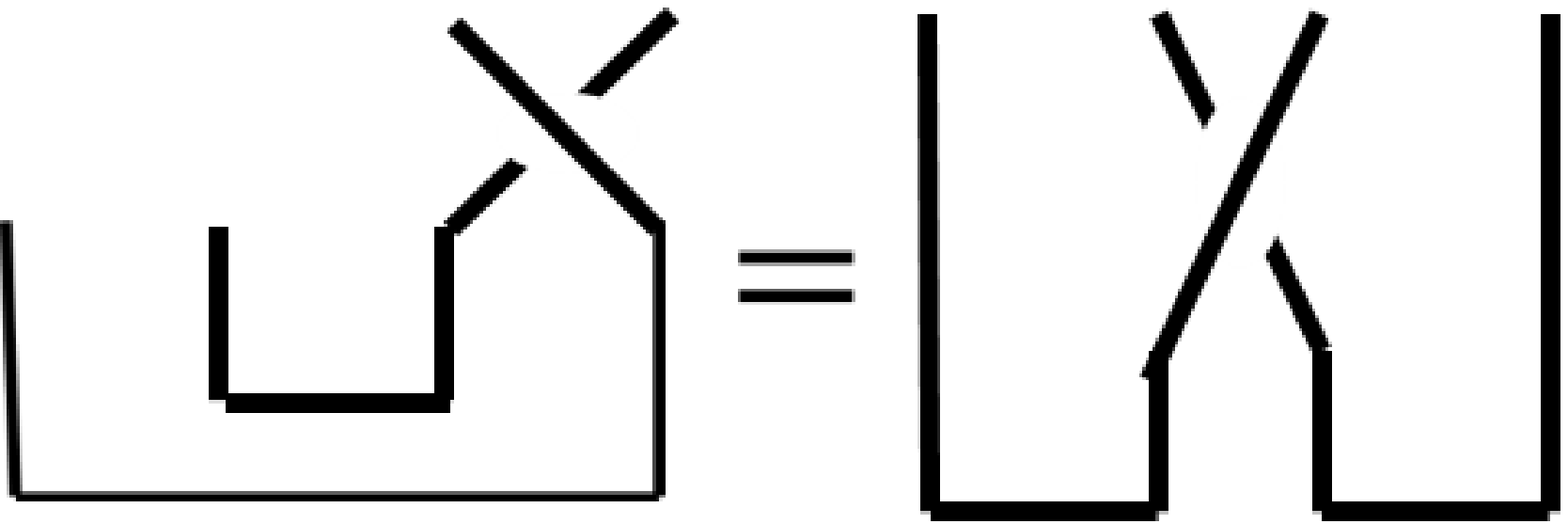}}}
\newcommand{\tb}{\raisebox{-0.3\height}{\includegraphics[height=0.6cm]{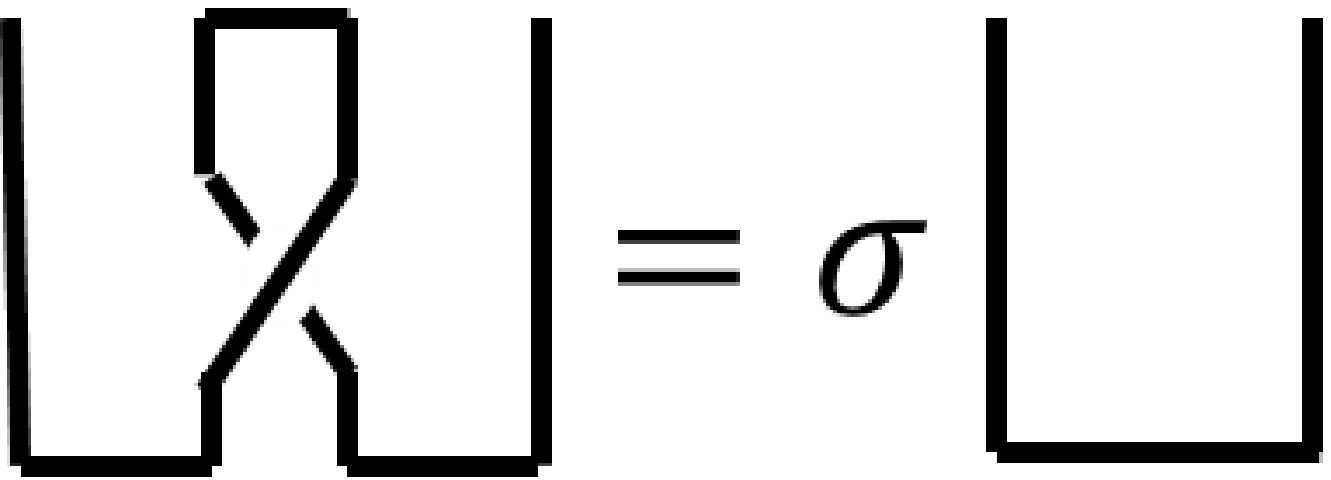}}}
\newcommand{\txx}{\raisebox{-0.3\height}{\includegraphics[height=0.6cm]{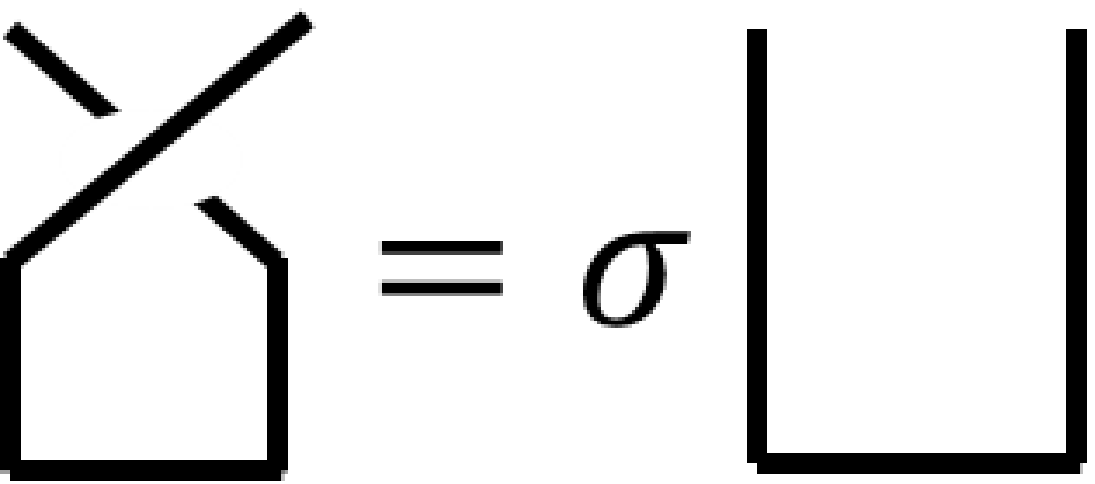}}}
\newcommand{\ty}{\raisebox{-0.3\height}{\includegraphics[height=0.6cm]{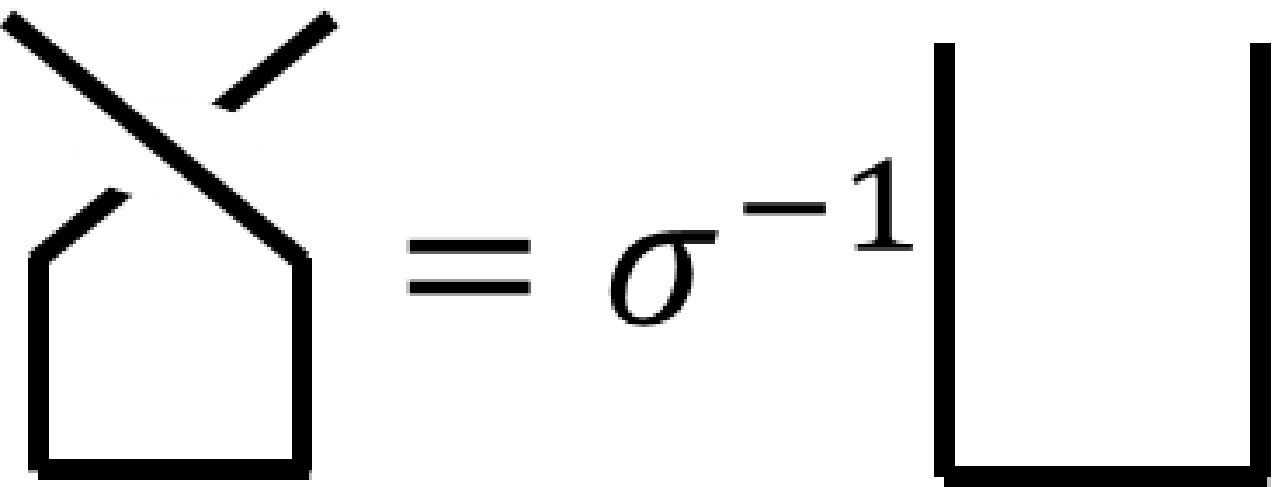}}}
\newcommand{\tz}{\raisebox{-0.3\height}{\includegraphics[height=0.6cm]{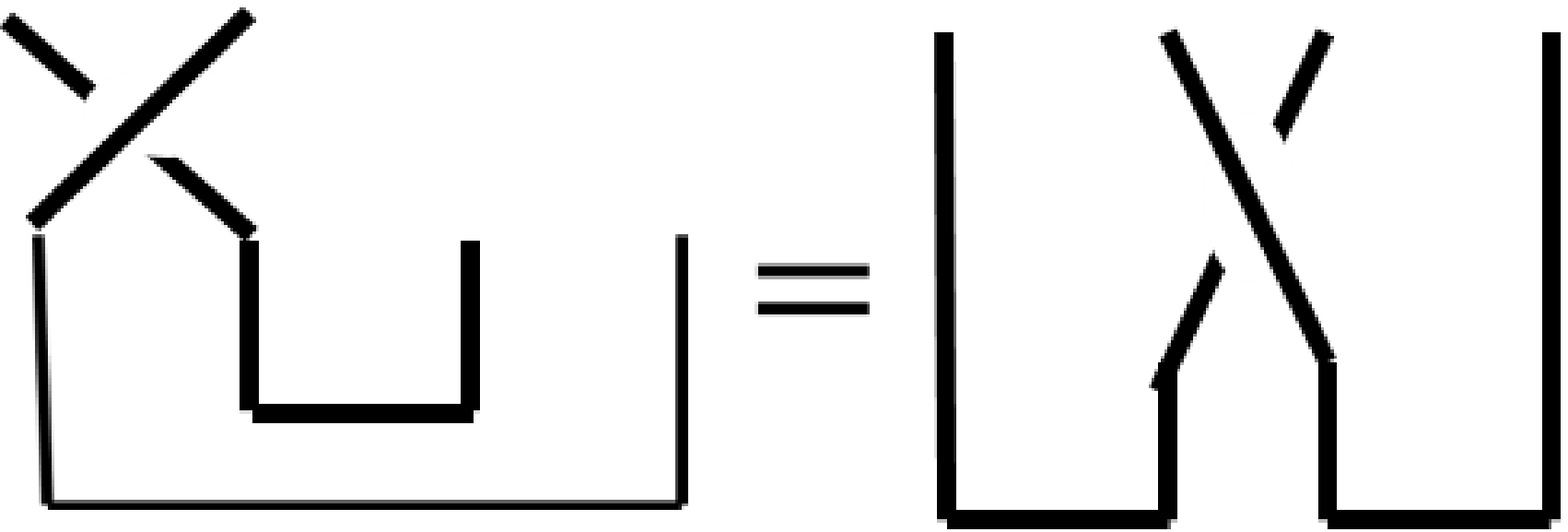}}}

\begin{document}

\title{Three-dimensional Topological Basis Associated with B-M-W algebra: Two Spin-1/2 Realization}

\author{Gangcheng Wang}
\author{Kang Xue}%
\email{xuekang@nenu.edu.cn}
\author{Chunfang Sun}
\author{Bo Liu}
\author{Yan Zhang}

\affiliation{%
School of Physics, Northeast Normal University, Changchun 130024,
People's Republic of China }

\date{\today}

\begin{abstract}
In this letter, we study the two-spin-1/2 realization for the Birman-Murakami-Wenzl (B-M-W) algebra and the corresponding Yang-Baxter $\breve{R}(\theta,\phi)$ matrix. Based on the two-spin-1/2 realization for the B-M-W algebra, the three-dimensional topological space, which is spanned by topological basis, is investigated. By means of such topological basis realization, the four-dimensional Yang-Baxter $\breve{R}(\theta,\phi)$ can be reduced to Wigner $D^{J}$ function with $J=1$. The entanglement and Berry phase in the spectral parameter space are also explored. The results show that one can obtain a set of entangled basis via Yang-Baxter $\breve{R}(\theta,\phi)$ matrix acting on the standard basis, and the entanglement degree is maximum when the $\breve{R}_{i}(\theta,\phi)$ turns to the braiding operator.
\end{abstract}

\pacs{02.10.Kn, 03.65.Ud, 02.40.-k}

\maketitle
\section{Introduction}\label{Sec:I}

The Yang-Baxter equation (YBE), which originated from solving the $\delta-$function interaction model by Yang in Ref. \cite{yang1967} and statistical model by Baxter in Ref. \cite{baxter1972}, plays an important role in statistical models, many-body problems and quantum integrable models, \emph{etc.} \cite{yang1994,kulish1982,kulish2010,jimbo1986,izergin1986}. Recently, braiding operator and YBE has been studied intensively due to their potential applications in the field of quantum information, quantum computation, topological quantum computation, \emph{etc}. In Ref. \cite{kauffman2004}, Kauffman and Lomonaco have explored the role of unitary of braiding operators in quantum computation theory. It is shown that the braiding matrices are universal quantum gates. Furthermore, based on Yang-Baxterization approach \cite{jones1991,cheng1991}, the quantum entanglement and Berry phase in spectral parameter space has been deeply investigated in Ref. \cite{chen2007}.

To investigate two-dimensional braiding statistical behavior under exchange of quasi-particles (such as Ising anyons), topological basis theory has been proposed in the Ref. \cite{freedman2002}. Such theory plays an important role in the field of topological quantum computation (TQC) \cite{kitaev2001,nayak2008} and motivates a novel way to study braid group realization, as well as YBE \cite{rowell2012,xue2012,ho2010}. Based on the development of topological quantum computation theory, Ge \emph{et al.} constructed the spin realization for the topological basis and reduced the four-dimensional solution of YBE to its two-dimensional form, then a linear-optical scheme has been proposed to simulate the Yang-Baxter system \cite{hu2008}. This scheme has recently been experimentally demonstrated successfully by Long \emph{et al.} \cite{zheng2013}, verifying both YBE and the velocity transformation relation predicted by the YBE given in Ref. \cite{hu2008}. Furthermore, the spin realizations for the topological basis are studied extensively. In Ref. \cite{niu2011}, the role of $\mathscr{L}_1-$norm in quantum mechanics has been studied through Wigner $D^{J}$ functions. The authors showed that the maximum of the $\mathscr{L}_1-$norm connectes with the maximally entangled states. It was to noting that the theory of topological basis provides an useful method to solve quantum spin chain models \cite{batchelor1992,batchelor2004,britta2010,ribeiro2013,sun2011}.

The braid matrix with two and three distinct eigenvalues related to the Temperley-Lieb (T-L) algebra and B-M-W algebra, respectively \cite{yang1994}. By means of topological basis associated with T-L algebra, one can reduce the solution of YBE associated with T-L algebra to Wigner $D^{1/2}$ function \cite{niu2011,benvegnu2006}. Then a question is raised naturally: Can we apply such approach to the three (or more than three) distinct eigenvalues case? The answer to this question is affirmative. In this paper, we will construct a two-spin-1/2 realization for B-M-W algebra, then the three dimensional topological basis is constructed. Consequently, the B-M-W algebra and its corresponding Yang-Baxter $\breve{R}(\theta,\phi)$ matrix are reduced to their three dimensional form, respectively. The Berry phase and quantum entanglement are also studied in details.

\section{Two-spin-1/2 realization of the B-M-W algebra}\label{sec:II}
The B-M-W algebra \cite{birman1989,wenzl1990,murakami1987}, denoted by $\mathcal {B}_{n}(\sigma,w)$, contains T-L algebra \cite{temperley1971} and braid algebra as its two sub-algebras. Let $E$ and $B$ be the T-L matrix and braid matrix, respectively. The notations $E_{i}$ and $B_{i}$ stand for $E_{i}\equiv I_{1}\otimes...\otimes I_{i-1}\otimes E\otimes I_{i+2}\otimes...\otimes I_{n}$ and $B_{i}\equiv I_{1}\otimes...\otimes I_{i-1}\otimes B\otimes I_{i+2}\otimes...\otimes I_{n}$, respectively. For each natural number $n$, the B-M-W algebra $\mathcal {B}_{n}(\sigma,w)$ is generated by $\{I,E_{i},B_{j},B^{-1}_{k};i,j,k=1,2,\cdots,n-1\}$ with the following algebraic relations
\begin{equation}\label{Eq:BMWA}
  \begin{array}{ll}
    B_{i}-B_{i}^{-1}=w (I-E_i) & 1\leq i\leq n; \\
    E_{i}B_i=B_{i}E_{i}=\sigma E_i & 1\leq i\leq n;\\
    E_{i}^2=d E_{i} & 1\leq i\leq n; \\
    E_{i}E_{i\pm 1}E_{i}=E_{i} & 2\leq i\leq n-1; \\
    B_{i}B_{i\pm 1}B_{i}=B_{i\pm 1}B_{i}B_{i\pm 1} & 2\leq i\leq n-1; \\
    B_{i\pm 1}B_{i}E_{i\pm 1}=E_{i}B_{i\pm 1}B_{i}=E_{i}E_{i\pm 1} & 2\leq i\leq n-1;\\
    B_{i\pm 1}E_{i}B_{i\pm 1}=B_{i}^{-1}E_{i\pm 1}B_{i}^{-1} & 2\leq i\leq n-1;\\
    B_{i\pm 1}E_{i}B_{i\pm 1}=B_{i}^{-1}E_{i\pm 1}B_{i}^{-1} &  2\leq i\leq n-1,
  \end{array}
\end{equation}
where the value of topological loop $d=1-(\sigma-\sigma^{-1})/w$. Let $\lambda_{1}$, $\lambda_{2}$ and $\lambda_{3}$ be three distinct eigenvalues of braid matrix. The second equation in Eqs. (\ref{Eq:BMWA}) implies that the topological parameter $\sigma$ is a eigenvalue of braid matrix. Without loss of generality, let the parameter $\sigma$ be the first eigenvalue of braid matrix (\emph{i.e.} $\sigma=\lambda_{1}$). The topological parameter $w$ can be cast summation of the other two eigenvalues of braid matrix (\emph{i.e.} $w=\lambda_{2}+\lambda_{3}$). One can verify that the other algebraic relations of B-M-W algebra in Ref. \cite{birman1989} can be derived from the Eqs. (\ref{Eq:BMWA}). In addition, the B-M-W algebra is easily understood in terms of knot diagrams in Refs. \cite{birman1989,wenzl1990,murakami1987}.

Let $\{\ket{ij};i,j=\ua,\da\}$ be the standard basis for the tensor product space $V^{1/2}\otimes V^{1/2}$. Direct calculation shows that the following matrices $E$ and $B$ satisfy B-M-W algebra
\begin{equation}\label{B-M-W_matrix}
\begin{array}{cc}
  E=\frac{1}{2}\left(
      \begin{array}{cccc}
        1 &i e^{-i\phi} &i e^{-i\phi} &  e^{-2i\phi} \\
        -i e^{i\phi} &  1 & 1 & -i e^{-i\phi} \\
        -i e^{i\phi} & 1 & 1 & -i e^{-i\phi} \\
         e^{2i\phi} & i e^{i\phi} &  i e^{i\phi} & 1 \\
      \end{array}
    \right),\\
    \\
  B=\frac{e^{i\frac{3\pi}{4}} }{2}\left(
                           \begin{array}{cccc}
                             1 & -e^{-i\phi} & -e^{-i\phi} & -e^{-2i\phi} \\
                             e^{i\phi} & 1 & -1 & e^{-i\phi} \\
                             e^{i\phi} & -1 & 1 & e^{-i\phi} \\
                             -e^{2i\phi} & -e^{i\phi} & -e^{i\phi} & 1 \\
                           \end{array}
                         \right),
\end{array}
\end{equation}
where $\phi$ is an arbitrary real parameter. The T-L matrix $E$ can be rewritten as the form of projection operator, $E=2\ket{\psi_{d}}\bra{\psi_{d}}$ with $\ket{\psi_{d}}=\frac{1}{2}(e^{-i\phi}\ket{\ua\ua}-i\ket{\ua\da}-i\ket{\da\ua}+e^{i\phi}\ket{\da\da})$. The quantum state $\ket{\psi_{d}}$ can be used to construct spin realization of the topological basis. Three distinct eigenvalues of braid matrix $B$ are $\lambda_{1}=e^{i5\pi/4}$, $\lambda_{2}=e^{i3\pi/4}$ (double-degenerate) and $\lambda_{3}=e^{i\pi/4}$.

For such special matrix representation of B-M-W algebra, the topological parameters $\sigma=\lambda_{1}=e^{i5\pi/4}$ and $w=\lambda_{2}+\lambda_{3}=\sqrt{2}i$. Then the topological loop value $d=1-(\sigma-\sigma^{-1})/w=2$. Let $s_{i}^{\pm}\equiv s_{i}^{1}\pm i s_{i}^{2}$ and $s_{i}^{3}$ be the $i$th spin operators. Then we realize the following spin-1 operator with two spin-1/2 operators
\begin{eqnarray}
  \begin{array}{cc}
    S^{\pm}_{ij}=-2(s_{i}^{\pm}s_{j}^{3}+s_{i}^{3}s_{j}^{\pm}), & S^{3}_{ij}=s_{i}^{3}+s_{j}^{3},
  \end{array}
\end{eqnarray}
where the spin-1 operators $S^{\pm}_{ij}\equiv S^{1}_{ij}\pm i S^{2}_{ij}$ and $S^{3}_{ij}$ satisfy the commutation relations $[S^{+}_{ij},S^{-}_{ij}]= 2 S^{3}_{ij}$ and $[S^{3}_{ij},S^{\pm}_{ij}]=\pm S^{\pm}_{ij}$. Let $$X_{i}=\frac{1}{2}(e^{-i\phi}S^{+}_{i,i+1}-e^{i\phi}S^{-}_{i,i+1})$$ and $$Y_{i}=2X^{2}_{i}+I.$$
Here the operator $X_{i}$ is anti-Hermitian (\emph{i.e.}, $X_{i}^{\dag}=-X_{i}$). In terms of $I$, $X_{i}$ and $Y_{i}$, the B-M-W algebra generators $E_{i}$ and $B_{i}$ can be recast as follows
\begin{equation}\label{EB-oper}
  \begin{array}{cc}
    E_{i}=\frac{1}{2}(I-2i X_{i}-Y_{i}), & B_{i}=\frac{1}{2} e^{i\frac{3}{4}\pi}(I+2X_{i}+Y_{i}),
  \end{array}
\end{equation}
which is a two-spin-1/2 realization for the B-M-W algebra. Consequently, we will study the physical applications of the Yang-Baxter system associated with such special B-M-W algebra realization.

\section{The $\breve{R}(\theta,\phi)$ matrix, entanglement and Berry phase}\label{Sec:III}
The solutions of YBE can be obtained via the so-called Yang-Baxterization approach of B-M-W algebra \cite{cheng1991}. In this section, we will introduce an alternative approach to Yang-Baxterize the B-M-W algebra. For the anti-Hermitian operator $X_{i}$, one can introduce a $\theta-$dependent unitary solution of YBE
\begin{equation}\label{R-operator}
  \breve{R}_{i}(\theta,\phi)=e^{\theta X_{i}}=\cos^{2}\frac{\theta}{2}(I+2\tan\frac{\theta}{2}X_{i}+\tan^{2}\frac{\theta}{2}Y_{i}).
\end{equation}
Here we used Taylor series to expand the exponential form. One can verify that the matrix $\breve{R}_{i}(\theta,\phi)$ is unitary (\emph{i.e.}, $[\breve{R}_{i}(\theta,\phi)]^{\dag}=[\breve{R}_{i}(\theta,\phi)]^{-1}=\breve{R}_{i}(-\theta,\phi)$). Its matrix form reads
\begin{equation}\label{R-matrix}
  \begin{array}{c}
     \breve{R}(\theta,\phi) =  \left(
                           \begin{array}{cccc}
                             \cos^{2}\frac{\theta}{2} & -\frac{1}{2}\sin\theta e^{-i\phi} & -\frac{1}{2}\sin\theta e^{-i\phi} & -\sin^{2}\frac{\theta}{2}e^{-2i\phi} \\
                             \frac{1}{2}\sin\theta e^{i\phi} & \cos^{2}\frac{\theta}{2} & -\sin^{2}\frac{\theta}{2} & \frac{1}{2}\sin\theta e^{-i\phi} \\
                             \frac{1}{2}\sin\theta e^{i\phi} & -\sin^{2}\frac{\theta}{2} & \cos^{2}\frac{\theta}{2} & \frac{1}{2}\sin\theta e^{-i\phi} \\
                             -\sin^{2}\frac{\theta}{2}e^{2i\phi} & -\frac{1}{2}\sin\theta e^{i\phi} & -\frac{1}{2}\sin\theta e^{i\phi} & \cos^{2}\frac{\theta}{2} \\
                           \end{array}
                         \right).
   \end{array}
\end{equation}
Direct calculation shows that the Yang-Baxter $\breve{R}_{i}(\theta,\phi)$-matrix satisfies the following YBE
\begin{equation}\label{YBE}
  \breve{R}_{i}(\theta_{1},\phi)\breve{R}_{i+1}(\theta_{2},\phi)\breve{R}_{i}(\theta_{3},\phi)=\breve{R}_{i+1}(\theta_{3},\phi)\breve{R}_{i}(\theta_{2},\phi)\breve{R}_{i+1}(\theta_{1},\phi),
\end{equation}
where the parameters $\theta_{1}$, $\theta_{2}$ and $\theta_{3}$ satisfy the following expressions
\begin{equation}\label{parameter-relation}
  \tan\frac{\theta_{2}}{2}=\frac{\tan\frac{\theta_{1}}{2}+\tan\frac{\theta_{3}}{2}}{1+\tan\frac{\theta_{1}}{2}\tan\frac{\theta_{3}}{2}}.
\end{equation}
By setting $v_{i}=\tan(\theta_{i}/2)$, this is just the additivity rule of Lorentz velocity: $v_{2}=(v_{1}+v_{3})/(1+v_{1}v_{3})$. Such velocity additivity rule are showed in Ref. \cite{xue2012}. We can verify the following expressions
\begin{equation*}
  \begin{array}{cc}
    \breve{R}_{i}(0,\phi)=I, & \breve{R}_{i}(\frac{\pi}{2},\phi)=e^{-i\frac{3}{4}\pi}B_{i}.
  \end{array}
\end{equation*}
Acting $\breve{R}_{i}(\theta,\phi)$ on the direct product state $\ket{kl}\equiv \ket{k}\otimes \ket{l}$ with $k,l=\ua,\da$, one can obtain an entangled basis
\begin{equation}\label{entangled-basis}
  \left(\begin{array}{c}
          \ket{\psi_{\ua\ua}} \\
          \ket{\psi_{\ua\da}} \\
          \ket{\psi_{\da\ua}} \\
          \ket{\psi_{\da\da}}
        \end{array}
  \right)_{i,i+1}=\breve{R}_{i}(\theta,\phi)\left(\begin{array}{c}
          \ket{\ua\ua} \\
          \ket{\ua\da} \\
          \ket{\da\ua} \\
          \ket{\da\da}
        \end{array}
  \right)_{i,i+1}.
\end{equation}
For a given pure state $\ket{\psi}=a\ket{\ua\ua}+b\ket{\ua\da}+c\ket{\da\ua}+d\ket{\da\da}$, the concurrence \cite{wooters1998} is $C(\ket{\psi})=2|ad-bc|$. Then we can obtain concurrence for the entangled basis. The four states in Eq. \ref{entangled-basis} possess the same degree of entanglement with $C(\ket{\psi_{kl}})=\sin^{2}\theta$. When $\theta=\pi/2$, the $\breve{R}_{i}(\theta,\phi)$ turns to the braiding operator and the entanglement degree for the four states is maximum.
Let $\phi$ be the time-dependent parameter with $\phi(t)=\omega t$.

The so-called Yang-Baxter Hamiltonian can be induced from the time-dependent gauge transformation
\begin{equation}\label{Hamiltonian}
  \hat{H}_{i}(\theta)=i\hbar \omega \left(\frac{\partial \breve{R}_{i}(\theta,\phi)}{\partial \phi}\right)\breve{R}_{i}^{\dag}(\theta,\phi).
\end{equation}
By introducing a new parameter $\vartheta=(\pi-\theta)/2$, the Yang-Baxter Hamiltonian can be rewritten as the following nuclear magnetic resonance (NMR) form
\begin{equation}\label{Hamiltonian-NMR}
  \hat{H}_{i}(\vartheta,\phi)=2\hbar \omega \cos\vartheta \vec{\textbf{n}}\cdot \vec{\textbf{S}},
\end{equation}
where $\vec{\textbf{n}}=(\sin\vartheta\cos\phi,\sin\vartheta\sin\phi,\cos\vartheta)$ is Bloch vector. The instantaneous eigenvalues for $H(\vartheta,\phi)$ are as follows
\begin{equation}
  \begin{array}{cc}
    E_{1,m_{s}}=2m_{s}\hbar \omega \cos\vartheta~(m_{s}=0,\pm 1),  & E_{0,0}=0.
  \end{array}
\end{equation}
The corresponding instantaneous eigen-states take as follows
\begin{equation}\label{eigenstates}
\begin{array}{l}
\ket{E_{1,1}} = \cos^{2}\frac{\vartheta}{2}e^{-2i\phi}\ket{\psi_{1,1}}-\frac{1}{\sqrt{2}}\sin\vartheta e^{-i\phi}\ket{\psi_{1,0}}-\sin^{2}\frac{\vartheta}{2}\ket{\psi_{1,-1}}, \\
\\
  \ket{E_{1,0}} =\frac{1}{\sqrt{2}}( \sin\vartheta e^{-i\phi}\ket{\psi_{1,1}}+\sqrt{2}\cos\vartheta\ket{\psi_{1,0}}+\sin\vartheta e^{i\phi}\ket{\psi_{1,-1}} ),\\
  \\
  \ket{E_{1,-1}} = \sin^{2}\frac{\vartheta}{2}\ket{\psi_{1,1}}+\frac{1}{\sqrt{2}}\sin\vartheta e^{i\phi}\ket{\psi_{1,0}}-\cos^{2}\frac{\vartheta}{2}e^{2i\phi}\ket{\psi_{1,-1}}, \\
  \\
  \ket{E_{0,0}} = \ket{\psi_{0,0}}.
\end{array}
\end{equation}
Here we used the following singlet-triplet basis
\begin{equation*}
  \begin{array}{l}
  \ket{\psi_{0,0}} = \frac{1}{\sqrt{2}}(\ket{\ua\da}-\ket{\da\ua}) \\
  \\
  \ket{\psi_{1,1}} = \ket{\ua\ua} \\
  \\
  \ket{\psi_{1,0}} = \frac{1}{\sqrt{2}}(\ket{\ua\da}+\ket{\da\ua}) \\
  \\
  \ket{\psi_{1,-1}} = \ket{\da\da}.
   \end{array}
\end{equation*}
According to Berry \cite{berry1984}, for a given instantaneous eigen-state $\ket{\psi}$, the so-called Berry phase reads $\gamma=i\int_{0}^{2\pi}\bra{\psi}\partial_{\phi}\ket{\psi}d\phi$. Substituting Eqs. (\ref{eigenstates}) into this formula, we obtain the corresponding Berry phase for this system
\begin{equation}
  \begin{array}{cc}
    \gamma_{1,m_{s}}=-m_{s}\Omega~(m_{s}=0,\pm 1),  & \gamma_{0,0}=0.
  \end{array}
\end{equation}
Here $\Omega=2\pi(1-\cos\vartheta)$ represents the solid angle of spectral parameter space. The Berry phase $\gamma_{1,m_{s}}$ and $\gamma_{0,0}$ correspond to the spin-1 and spin-0 subspace, respectively. That is to say the Yang-Baxter system can be decomposed into two subspaces $V^{1}$ and $V^{0}$ (\emph{i.e.}, $V^{1/2}\otimes V^{1/2}=V^{1}\oplus V^{0}$).

\section{Three-dimensional topological space realization}\label{Sec:IV}
For the B-M-W algebra, the graphic operator forms of $E_{i}$, $B_{i}$ and $B_{i}^{-1}$ read \cite{xue2012}
\begin{equation}\label{graphic-operator}
  \begin{array}{ccc}
    E_{i}=\Ei, & B_{i}=\Bi, & B_{i}^{-1}=\Bim.
  \end{array}
\end{equation}
The topological basis for the B-M-W algebra is constructed by the following three graphic states
 \begin{equation*}
   \begin{array}{ccc}
     \gx, & \gy, & \gz.
   \end{array}
 \end{equation*}
For the braid matrix with two distinct eigenvalues, the third graphic states can be recast as the liner superposition of the first two graphic states. For this special realization, the graphic states can be realized by the four spin-1/2 states as following
\begin{equation*}
    \begin{array}{l}
  \gx = 2\ket{\psi_{d}}_{12}\ket{\psi_{d}}_{34},  \\
  \\
  \gy = 2\ket{\psi_{d}}_{14}\ket{\psi_{d}}_{23},  \\
  \\
  \gz = 2B_{2}\ket{\psi_{d}}_{12}\ket{\psi_{d}}_{34},
    \end{array}
\end{equation*}
where $\ket{\psi_{d}}_{ij}=\frac{1}{2}(e^{-i\phi}\ket{\ua\ua}_{ij}-i\ket{\ua\da}_{ij}-i\ket{\da\ua}_{ij}+e^{i\phi}\ket{\da\da}_{ij})$. The topological space for the B-M-W algebra is spanned by the following three orthogonal graphic states
\begin{equation}\label{basis}
\begin{array}{l}
\ket{e_{1}} = \frac{1}{2}\gx, \\
\\
  \ket{e_{2}} = \frac{1}{2}\left(\gx-\gy-e^{i\frac{5}{4}\pi}\gz\right), \\
  \\
  \ket{e_{3}} = \frac{1}{2}\left(\gx+\sqrt{2}e^{i\frac{3}{4}\pi}\gy+ \sqrt{2}e^{i\frac{1}{2}\pi}\gz\right).
\end{array}
\end{equation}
The algebraic relations in Eqs. (\ref{Eq:BMWA}) imply that the following topological relations
\begin{eqnarray}\label{topo-equivalence}
\begin{array}{lll}
  \txx, & \ty,  & \tz, \\
  \\
  \ta, & \tb, & \loopa=d^{2},\\
  \\
  \loopx=\loopxt=d, & \loopy=\sigma^{-1} d, & \loopz=\sigma d.
\end{array}
\end{eqnarray}
Acting $E_{i}$ and $B_{i}$ in Eq. (\ref{graphic-operator}) on topological basis, we obtain the corresponding $3\times 3$ representation. We denote the reduced form of $B_{1}$, $B_{2}$, $E_{1}$ and $E_{2}$ by $\mathscr{A}$, $\mathscr{B}$, $E_{A}$ and $E_{B}$, respectively. The matrix elements of the reduced representations are defined as $\mathscr{A}_{ij}=\bra{e_{i}}B_{1}\ket{e_{j}}$, $\mathscr{B}_{ij}=\bra{e_{i}}B_{2}\ket{e_{j}}$, $(E_{A})_{ij}=\bra{e_{i}}E_{1}\ket{e_{j}}$, and $(E_{B})_{ij}=\bra{e_{i}}E_{2}\ket{e_{j}}$. By using the topological basis in Eqs. (\ref{basis}) and relations of topological equivalence in Eqs. ({\ref{topo-equivalence}}), we obtain the following reduced matrix representation of B-M-W algebra
\begin{eqnarray*}
\begin{array}{c}
  \mathscr{A}=e^{i\frac{3}{4}\pi}\left(
                           \begin{array}{ccc}
                             i & 0 & 0 \\
                             0 & 1 & 0 \\
                             0 & 0 & -i \\
                           \end{array}
                         \right), \quad E_{A}=\left(
                           \begin{array}{ccc}
                             2 & 0 & 0 \\
                             0 & 0 & 0 \\
                             0 & 0 & 0 \\
                           \end{array}
                         \right),\\
                         \\
  E_{B} = \left(
                           \begin{array}{ccc}
                             \frac{1}{2} & \frac{1}{\sqrt{2}}i e^{-i\frac{\pi}{4}} & -\frac{1}{2}e^{-i\frac{\pi}{2}} \\
                             -\frac{1}{\sqrt{2}}i e^{i\frac{\pi}{4}} & 1 & \frac{1}{\sqrt{2}}i e^{-i\frac{\pi}{4}} \\
                             -\frac{1}{2}e^{i\frac{\pi}{2}} & -\frac{1}{\sqrt{2}}i e^{i\frac{\pi}{4}} & \frac{1}{2} \\
                           \end{array}
                         \right),\\
                         \\
  \mathscr{B}= e^{i\frac{3}{4}\pi}\left(
                           \begin{array}{ccc}
                             \frac{1}{2} & -\frac{1}{\sqrt{2}}e^{-i\frac{\pi}{4}} & \frac{1}{2}e^{-i\frac{\pi}{2}} \\
                             \frac{1}{\sqrt{2}}e^{i\frac{\pi}{4}} & 0 & -\frac{1}{\sqrt{2}}e^{-i\frac{\pi}{4}} \\
                             \frac{1}{2}e^{i\frac{\pi}{2}} & \frac{1}{\sqrt{2}}e^{i\frac{\pi}{4}} & \frac{1}{2} \\
                           \end{array}
                         \right).
\end{array}
\end{eqnarray*}
For the following convenience, we introduce a $U(1)$ transformation matrix $u(\phi)=e^{i(\frac{\pi}{4}-\phi)}\ket{e_{1}}\bra{e_{1}}+e^{-i(\frac{\pi}{4}-\phi)}\ket{e_{3}}\bra{e_{3}}$. The matrices $\mathscr{B}$ and $E_{B}$ can be recast as follows
\begin{equation*}
  \begin{array}{l}
    \mathscr{B}'=u(\phi)\mathscr{B} u(\phi)^{\dag}=e^{i\frac{3}{4}\pi}\left(
                           \begin{array}{ccc}
                             \frac{1}{2} & -\frac{1}{\sqrt{2}}e^{-i\phi} & \frac{1}{2}e^{-2i\phi} \\
                             \frac{1}{\sqrt{2}}e^{i\phi} & 0 & -\frac{1}{\sqrt{2}}e^{-i\phi} \\
                             \frac{1}{2}e^{2i\phi} & \frac{1}{\sqrt{2}}e^{i\phi} & \frac{1}{2} \\
                           \end{array}
                         \right),\\
                         \\
      E'_{B}=u(\phi)E_{B} u(\phi)^{\dag}=\left(
                           \begin{array}{ccc}
                             \frac{1}{2} & \frac{1}{\sqrt{2}}i e^{-i\phi} & -\frac{1}{2}e^{-2i\phi} \\
                             -\frac{1}{\sqrt{2}}i e^{i\phi} & 1 & \frac{1}{\sqrt{2}}i e^{i\phi} \\
                             -\frac{1}{2}e^{2i\phi} & -\frac{1}{\sqrt{2}}i e^{i\phi} & \frac{1}{2} \\
                           \end{array}
                         \right).
  \end{array}
\end{equation*}
Then we can verify that $\mathscr{A}$, $\mathscr{B}'$, $E_{A}$ and $E'_{B}$ satisfy the reduce three-dimensional B-M-W algebra relations. In topological space, spin operators can be written as
\begin{equation*}
  \begin{array}{l}
     S_{T}^{+}=-\sqrt{2}(\ket{e_{1}}\bra{e_{2}}+\ket{e_{2}}\bra{e_{3}}),\\
     \\
     S_{T}^{+}=-\sqrt{2}(\ket{e_{2}}\bra{e_{1}}+\ket{e_{3}}\bra{e_{2}}),\\
     \\
     S_{T}^{3}=\ket{e_{1}}\bra{e_{1}}-\ket{e_{3}}\bra{e_{3}}.
   \end{array}
\end{equation*}
The matrices $X_{T}$ and $Y_{T}$ read
\begin{equation*}
  \begin{array}{cc}
     X_{T}=\frac{1}{2}(e^{-i\phi}S_{T}^{+}-e^{i\phi}S_{T}^{-})
     ,\quad & Y_{T}=2X_{T}^{2}+I_{T}.
   \end{array}
\end{equation*}
Here the notation $I_{T}$ stands for Identity matrix for the topological space. In terms of $X_{T}$ and $Y_{T}$, The matrices $\mathscr{B}'$ and $E'_{B}$ can be recast as follows
\begin{equation}
  \begin{array}{cc}
    E'_{B} = \frac{1}{2}(I-2i X_{T}-Y_{T}), & \mathscr{B}' = \frac{1}{2}e^{i\frac{3}{4}\pi}(I+2X_{T}+Y_{T}).
  \end{array}
\end{equation}
In topological subspace, Yang-Baxter $\breve{R}(\theta,\phi)$ matrix can be reduced to the following $3\times 3$ form
\begin{equation}\label{label{3D-representation}}
  \begin{array}{c}
    \mathscr{A}(\theta)=e^{i\theta S_{T}^{3}},\\
    \\
    \mathscr{B}(\theta,\phi)=e^{\theta X_{T}}=\cos^{2}\frac{\theta}{2}(I+2\tan\frac{\theta}{2}X_{T}+\tan^{2}\frac{\theta}{2}Y_{T}).
  \end{array}
\end{equation}
Their matrix forms read
\begin{equation}
\begin{array}{c}
  \mathscr{A}(\theta)=\left(
                               \begin{array}{ccc}
                                 e^{i\theta} & 0 & 0 \\
                                 0 & 1 & 0 \\
                                 0 & 0 & e^{-i\theta} \\
                               \end{array}
                             \right), \\
                             \\
  \mathscr{B}(\theta,\phi)=\left(
                               \begin{array}{ccc}
                                 \cos^{2}\frac{\theta}{2} & -\frac{1}{\sqrt{2}}\sin\theta e^{-i\phi} & \sin^{2}\frac{\theta}{2} e^{-2i\phi} \\
                                 \frac{1}{\sqrt{2}}\sin\theta e^{i\phi} & \cos\theta & -\frac{1}{\sqrt{2}}\sin\theta e^{-i\phi} \\
                                 \sin^{2}\frac{\theta}{2} e^{2i\phi} & \frac{1}{\sqrt{2}}\sin\theta e^{i\phi} & \cos^{2}\frac{\theta}{2} \\
                               \end{array}
                             \right).
\end{array}
\end{equation}
The matrices $\mathscr{A}(\theta)$ and $\mathscr{B}(\theta,\phi)$ satisfy the following reduced YBE
\begin{equation}\label{rYBE}
  \mathscr{A}(\theta_{1})\mathscr{B}(\theta_{2},\phi)\mathscr{A}(\theta_{3})=\mathscr{B}(\theta_{3},\phi)\mathscr{A}(\theta_{2})\mathscr{B}(\theta_{1},\phi),
\end{equation}
where the parameters $\theta_{1}$, $\theta_{2}$ and $\theta_{3}$ satisfy the relation given in Eq. (\ref{parameter-relation}). The matrix $\mathscr{B}(\theta,\phi)$ is Wigner $D^{J}$ function with $J=1$ and the matrix $\mathscr{A}(\theta)$ is its diagonal form. When the parameter $\theta=\pi/2$, the $\mathscr{B}(\theta,\phi)$ matrix is reduced to $\mathscr{B}'$ matrix. And the Hamiltonian in Eq. (\ref{Hamiltonian}) is reduced to its three-dimensional form $H_{T}(\theta)=i\hbar \omega \left(\frac{\partial \mathscr{B}(\theta,\phi)}{\partial \phi}\right)\mathscr{B}^{\dag}(\theta,\phi)$. By introducing a new parameter $\vartheta=(\pi-\theta)/2$, the reduced Yang-Baxter Hamiltonian can be rewritten as the following nuclear magnetic resonance (NMR) form $\hat{H}_{T}(\vartheta,\phi)=2\hbar \omega \cos\vartheta \vec{\textbf{n}}\cdot \vec{\textbf{S}}_{T}$. This shows that the Hamiltonian reduced to its three-dimensional form by means of the topological basis associated with the B-M-W algebra.

\section{Summary and discussion}\label{Sec:V}
In this paper, we studied two-spin-1/2 realization for the B-M-W algebra and the corresponding Yang-Baxter $\breve{R}(\theta,\phi)$ matrix. The topological basis associated with the two-spin-1/2 realization for the B-M-W algebra has been also investigated. By means of such topological basis realization, the four-dimensional Yang-Baxter $\breve{R}(\theta,\phi)$ has be reduced to Wigner $D^{J}$ function with $J=1$. The entanglement and Berry phase in the spectral parameter space have been also studied in this paper. We has been obtained a set of entangled basis via Yang-Baxter $\breve{R}(\theta,\phi)$ matrix acting on the standard basis, and the entanglement degree is maximum when the $\breve{R}_{i}(\theta,\phi)$ turns to the braiding operator.

Let us make some discussions to end this paper. ($i$) In fact, we can construct spin chain model of B-M-W type with the B-M-W algebraic generators, and the approach of topological basis will play an important role in the ground state solving. ($ii$) In Ref. \cite{hu2008}, the authors simulated Yang-Baxter system by means of linear quantum optics. In the current paper, three-dimensional topological basis is introduced to reduce the B-M-W algebra and associated YBE. We think such YBE system can be simulated with NMR system \cite{aucc2013}. ($iii$) In Ref. \cite{niu2011}, the authors showed that the Wigner $D^{J}$ function satisfies the so-called reduced YBE, and they studied the braiding matrix and the corresponding solution of YBE for the case of $J=1/2$. In the current paper, we studied the case of $J=1$, and there are $m$ ($m=2J+1$) distinct eigenvalues for the braiding matrix. We conjecture that $m-$dimensional ($m\geq 4$) topological space can be constructed by means of matrix representation of B-M-W algebra with $m$ distinct eigenvalues. Such matrix representations, as well as the reduced matrix representations, have potential applications in the fields of quantum computation and topological quantum computation. This work will encourage us to study the relations between quantum computation and topological quantum computation for more higher dimension.

\section*{Acknowledgments}
We thank Guilu Long, Shuangwei Hu, and Zhihai Wang for helpful comments. This work was supported by the NSF of China (Grant No. 11175043, No. 11247005 and No. 11205028), the Plan for Scientific and Technological Development of Jilin Province (No. 20130522145JH). C. F. Sun was also supported in part by the Government of China through CSC.

\end{document}